# PPSpeech: Phrase based Parallel End-to-End TTS System

*Yahuan Cong[1], Ran Zhang[2], Jian Luan[2]*

[1]Beijing University of Posts and Telecommunications, China
[2]Xiaoice, Software Technology Center, Microsoft
congyahuan@bupt.edu.cn, {Ran.Zhang, jianluan}@microsoft.com

## Abstract

Current end-to-end autoregressive TTS systems (e.g. Tacotron 2) have outperformed traditional parallel approaches on the quality of synthesized speech. However, they introduce new problems at the same time. Due to the autoregressive nature, the time cost of inference has to be proportional to the length of text, which pose a great challenge for online serving. On the other hand, the style of synthetic speech becomes unstable and may change obviously among sentences. In this paper, we propose a Phrase based Parallel End-to-End TTS System (PPSpeech) to address these issues. PPSpeech uses autoregression approach within a phrase and executes parallel strategies for different phrases. By this method, we can achieve both high quality and high efficiency. In addition, we propose acoustic embedding and text context embedding as the conditions of encoder to keep successive and prevent from abrupt style or timbre change. Experiments show that, the synthesis speed of PPSpeech is much faster than sentence level autoregressive Tacotron 2 when a sentence has more than 5 phrases. The speed advantage increases with the growth of sentence length. Subjective experiments show that the proposed system with acoustic embedding and context embedding as conditions can make the style transition across sentences gradient and natural, defeating Global Style Token (GST) obviously in MOS.

**Index Terms**: phrase level, end-to-end speech synthesis, acoustic embedding, text context embedding

## 1. Introduction

Deep neural network based systems have become more and more popular for TTS (text-to-speech) task, such as Tacotron [1][2] Deep Voice [3]. Those models usually generate Mel-spectrogram autoregressively from text input at sentence level and then synthesize speech from the Mel-spectrogram using vocoder such as WaveNet [4] or WaveGlow [5]. Compared with non-autoregressive sequence generation, such as concatenative [6] and parametric [7], autoregressively neural networks significantly improve the quality of synthesized speech. However, all such models generate Mel-spectrogram conditioned on the previously generated ones [9]. Due to the autoregressive nature, they suffer from slow inference speed and lack of controllability (voice speed or prosody control). Real-time and high-quality speech synthesis remains a challenging task [5].

To speed up Mel-spectrogram generation, FastSpeech [9] adopts a feed-forward network based on the self-attention in Transformer [8] to generates Mel-spectrograms parallelly but still has a slight quality gap against the autoregressive method. Other synthetic acceleration methods [10][11][12] are also proposed from the perspective of replacing the autoregressive model, but they are hard to train and implement. More importantly, the quality gap somewhat remains.

In this paper, we propose a Phrase based Parallel End-to-End TTS System (PPSpeech). It implements speech synthesis parallelly in phrase-level, which greatly shortens the time in comparison to sentence-level autoregressive speech synthesis. Each phrase is still generated in an autoregressive way to keep the high naturalness. In addition, acoustic embedding and text context embedding are employed to make the synthesized speech more natural and expressive without abrupt timbre and style change. Although GST [13] reported the style control through acoustic embedding, we find the text context also plays an important role and need to considered.

The rest of paper is organized as follows. Section 2 elaborates model architecture. The experimental results are presented in section 3 and conclusions are drawn in section 4.

## 2. Model Architecture

PPSpeech system (see Figure 1) consists of three components: text processor, conditional Tacotron 2 network and WaveGlow. The text processor analyzes the input text to get phoneme sequence and segment sentence into phrases. Then phoneme sequence is sent to conditional Tacotron 2 network as input while extracted acoustic feature and phrase context are embedded as the condition of encoder. Mel-spectrograms are generated as the output of network and passed to WaveGlow vocoder. Finally, time-domain waveform samples are obtained.

### 2.1. Text processor

Text processor consists of a phrase boundary detector and a grapheme-to-phoneme (G2P) convertor. Phrase boundary detector is used to predict intonation phrase boundaries(L3) in the input sentence. After the prediction, we separate the sentence into batches of phrases. In PPSpeech, an expanded CRF supporting dynamic features [14] is used for boundary prediction. The feature template we used is listed as follows:

- Text
- POS
- Number of syllables in the word
- Is the word followed by punctuation
- Above features' combination
- Bigram feature
- Number of words from/to punctuation
- Number of syllables from/to punctuation
- Number of words from previous L3

G2P convertor segments input phrase text into words and transcribes each word into phonemes. After the G2P module, batches of phrases are converted to batches of phone sequences, and then fed into the conditional Tacotron2 network.

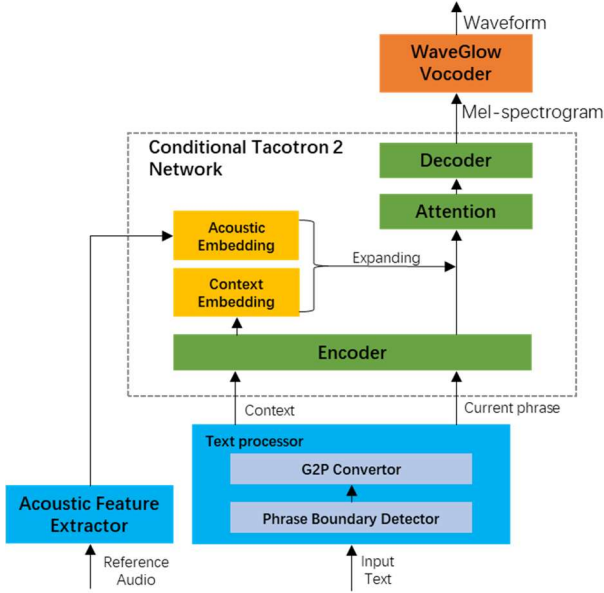

Figure 1: *An architecture of PPSpeech.*

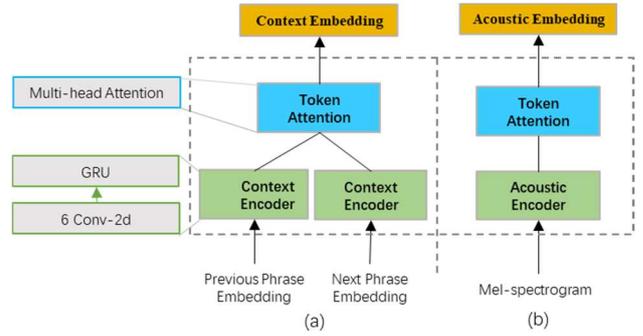

Figure 2: *Structure diagram of condition embeddings. (a) context embedding, (b) acoustic embedding.*

**2.2. Conditional Tacotron 2 network**

Standard Tacotron 2 network consists of an encoder and a decoder with attention. The encoder converts phoneme sequence into latent feature representation which the decoder consumes to predict Mel-spectrogram. It is noted that only input phoneme sequence of current phrase cannot guarantee the coherence across phrases. Therefore, two extra conditions are concatenated into the encoder to better control it.

*2.2.1. Context embedding*

In PPSpeech, the phoneme sequence of a phrase rather than whole sentence is used as input. It means the attention is only implemented inside the phrase and not across phrases. In addition, the phrase at different positions of sentence, should have different intonation and stress etc. However, Tacotron 2 cannot model it well [9]. The shorter the phrases are, the more obvious inconsistency among phrases will be. To keep the prosody of neighboring phrases cohesive in a sentence and even smooth across sentences, context information is necessary for acoustic modeling. Therefore, a context embedding network is designed in our phrase-level speech synthesis system.

For context embedding, the phoneme sequence of the previous phrase and the next phrase will pass through the encoder as the same as current phrase did (see Figure 1). Then, encoder outputs of them are send to context encoder, respectively. As shown in Figure 2(a), the context encoder is a stack of six 2-D convolutional layers followed by a GRU layer. All the convolutional layers are 3×3 kernel, 2×2 stride, with batch normalization and ReLU activation function. 32, 32, 64, 64, 128 and 128 filters are applied for the six convolutional layers, respectively. The sequence length is variable, since the previous and the next phrase may have different number of phonemes. A 128-width gated recurrent unit (GRU) layer is then applied to map the variable sequence length into one. Finally, 128-d vector is output after a fully connected layer followed with softmax activation function. It should be noted that the context encoders for the previous and the next phrase share the same parameters.

The context encoder outputs are concatenated and passed to a token attention layer, where it is used as the query of the attention. A table of tokens act as both keys and values. The attention is to get a weighted sum of tokens according to the similarity of query to keys. The token table is randomly initialized and shared across all the training data. We use the same settings as the style token layer in GST [13], with 10 initialized embeddings, and the initialized embeddings are 256 dimensions.

*2.2.2. Acoustic embedding*

For high quality TTS voice, a large recording corpus are often required. The voice talent can hardly finish them at one recording session. Therefore, the strict consistency of timber and speaking style among sessions are difficult to guarantee. Particularly, the recordings data are required to be rich in expression, such as audio book scenario. It usually results in unstableness issue in synthesized speech. To handle it, in PPSpeech, an acoustic embedding is employed as another condition to control speaking style.

The structure of acoustic embedding is shown in Figure 2(b), which includes two blocks of acoustic encoder and token attention. The acoustic encoder and token attention have the same network structure with context encoder and token attention respectively in context embedding. However, their model parameters are trained independently and its output dimension 128. Mel-spectrograms of current phrase are sent to acoustic encoder in training phase. While in inferring phase, a reference audio is used to generate the acoustic embedding. Generally, the same reference audio is applied for the phrases in one sentence to keep the speaking style stable.

*2.2.3. Condition*

Acoustic embedding and context embedding outputs are concatenated as the condition of Tacotron 2 network. Since their sequence length becomes one, they need to be expanded by duplication to the sequence length of encoded current phrase before concatenation. Specifically, 128-d acoustic embedding, 256-d context embedding, and 512-d encoder output are concatenated into an 896-d vector. Then it is passed to the following attention and decoder.

**2.3. WaveGlow vocoder**

WaveGlow is a NN-based vocoder that produces audio by sampling from a distribution. Although its performance is slightly worse than WaveNet, it works with high efficiency due to the nature of non-autoregressive. For this reason, we select

WaveGlow as the vocoder of PPSpeech to generate audio from predicted Mel-spectrogram.

## 3. Experiments

### 3.1. Training Setup

Our model is trained on an NVIDIA P100 GPU. We use a learning rate of $1 \times 10^{-3}$, exponentially decaying to $5 \times 10^{-5}$. For every 10 epochs, the learning rate drops by 0.95.

#### 3.1.1. Data corpus

All the experiments are conducted on an 8-hour Mandarin audiobook dataset collected from a male voice talent. Among them, 7554 sentences are randomly selected out for training and the rest 100 for test. 80-d Mel-spectrograms are extracted with 50ms window at every 12.5ms. In our data, each sentence is divided into about 3.7 phrases, and each phrase contains approximately 15.4 phonemes. The longest sentence has 42 phrases, while the shortest one has only 1 phrase.

#### 3.1.2. Sliding text window

Comparing with sentence level system, phrase level system suffers from much less context information, which brings more challenges to prosody modeling. To overcome this issue, we introduce a sliding window strategy which incorporates multiple phrases for training. K phrases are input as current phrase. Besides, the previous M phrases and the next N phrases will be used as reference information for condition embeddings.

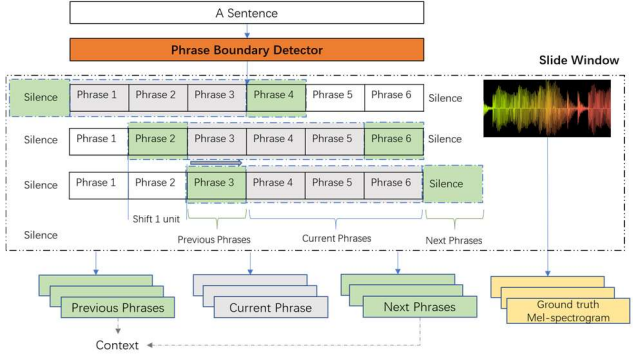

Figure 3: *Illustration of sliding text window. Every unit is a phrase divided by a phrase boundary. M=1, N=1 and K=3, shift = 1 are applied.*

In the experiments, M and N are always set to be 1. K=3 is used for training to enhance the model robustness. As shown in Figure 3, through sliding text window, three inputs can be obtained from the sentence. While in inference phase, we set k = 1 to accelerate the synthesis speed so that every phrase can be generated in parallel.

### 3.2. Evaluation

#### 3.2.1. Phrase-level TTS systems

There are three phrase-level TTS systems are realized for comparison in a Mean Opinion Score (MOS) test:

- Phrase-Level-Tacotron2, without condition embeddings.

---

[1] Samples available at https://yahcong.github.io/PPSpeech/

- Phrase-Level-Tacotron2-GST, a phrase level Tacotron 2 TTS system with GST condition. Mel-spectrogram is the input of GST. The hyperparameters of GST are set according to [13]
- PPSpeech, the proposed system, condition on both acoustic embedding and context embedding.

100 sentences in the test set are synthesized by each system. Each sample is rated by at least 10 listeners on a scale from 1 to 5. The results given in Table 1 show that GST can help to improve the speaking style stableness of phrase-level synthesis. However, PPSpeech with context embedding can further enhance the prosody coherence and defeat Phrase-Level-Tacotron2 and Phrase-Level-Tacotron2-GST remarkably. [1]

Table 1: *MOS for Phrase-Level-Tacotron2, Phrase-Level-Tacotron2-GST and PPSpeech.*

| System Name | score |
|---|---|
| Phrase-Level-Tacotron2 | 3.665 ± 0.125 |
| Phrase-Level-Tacotron2-GST | 3.823 ± 0.121 |
| **PPSpeech** | 3.955 ± 0.122 |
| Recording | 4.148 ± 0.100 |

To further validate the effect of condition embedding, Mel-spectrograms of one example are drawn in Figure 4. Herein, (a) is synthesized by Phrase-Level-Tacotron2, (b) is by PPSpeech and (c) is recoding. As we can see, the Mel-spectrogram of (b) is obviously more similar to the recording, especially the pattern in the white square box. Besides, the speaking rate of (b) is also closer to the recoding than (a). They both approve that our condition embeddings can effectively reconstruct the prosody and speaking style well and produce more natural synthesis speech.

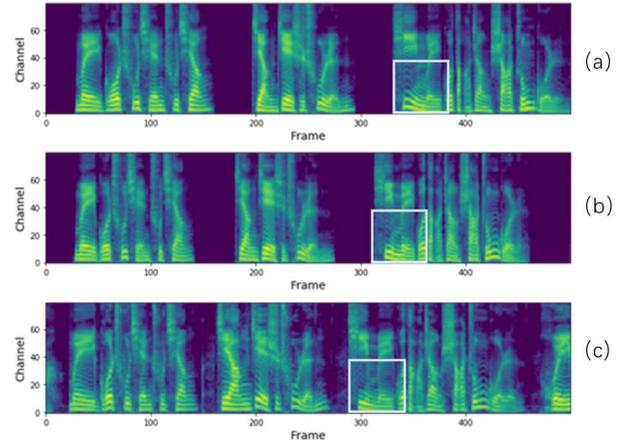

Figure 4: *An example of (a) Phrase-Level-Tacotron2, (b) PPSpeech, (c) Recording.*

In our condition embedding, there are two components: acoustic embedding and context embedding. GST [13] has proved the effect of acoustic embedding. Then we compare PPSpeech to Phrase-Level-Tacotron2-GST in Figure 5 to validate the advantage of context embedding. The Mel-spectrogram in the picture comes from a sentence which has three phrases, and all of them have very similar contents. (a) is generated by Phrase-Level-Tacotron2-GST while (b) is generated by PPSpeech. It can be found that every phrase in (a) has similar expression, regardless of their position or context in the sentence. However, the phrases in (b) have more variation depending on their position and context. Overall, (b) seems more consistent with (c).

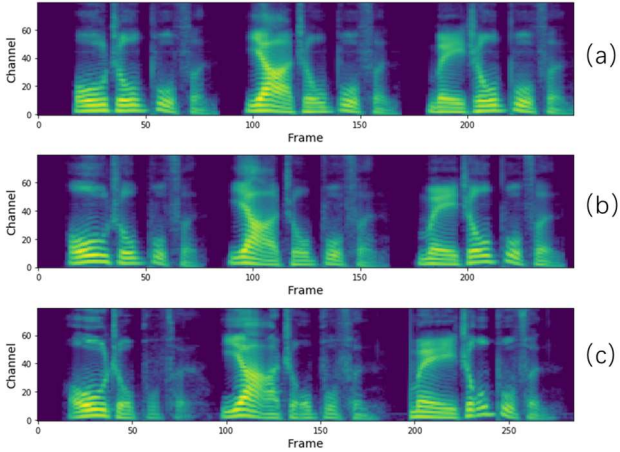

Figure 5: *An example for Phrase-Level-Tacotron2-GST and PPSpeech. (a) GST, (b) PPSpeech, (c) recording.*

### 3.2.2. Tacotron 2 vs. PPSpeech

Above experiments compare the different phrase level TTS systems and PPSpeech show the best performance. However, there are still concerns about the quality gap of PPSpeech to sentence level Tactoron 2 and how much time can be saved by PPSpeech.

#### 3.2.2.1 Overall quality

A side-by-side evaluation is conducted to compare the audio quality between PPSpeech and Sentence-Level-Tacotron2. For each pair of utterances, listeners are asked to give a score ranging from -3 (PPSpeech is much worse than Sentence-Level-Tacotron2) to 3 (PPSpeech is much better than Sentence-Level-Tacotron2).

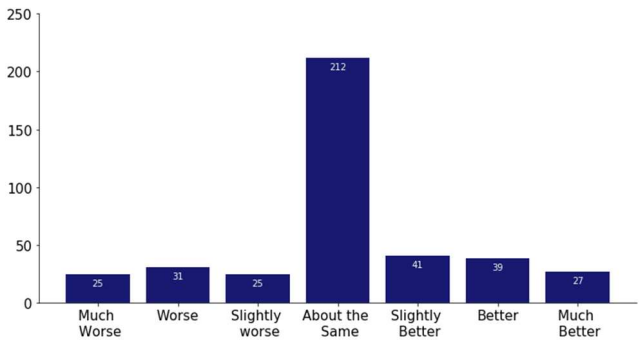

Figure 6: *PPSpeech vs. Sentence-Level-Tacotron2.*

Results are shown in Figure 6 for detailed breakdown. 53.0% judges are "About the same" while the overall mean score is +0.095. It indicates that although speech is synthesized in parallel at the phrase level, PPSpeech can achieve comparable or even slight better overall quality than Sentence-Level-Tacotron2. It must be attributed to the acoustic embedding and text context embedding.

#### 3.2.2.2 Inference speed

To evaluate the inference speed change, we run these two systems to generate batches of test sentences on a server with NVIDIA P100 GPU. The sentences are grouped by the number of phrases they contain. There are eight groups, phrase number ranges from 5 to 40 at a step of 5. For each group, 30 sentences are prepared.

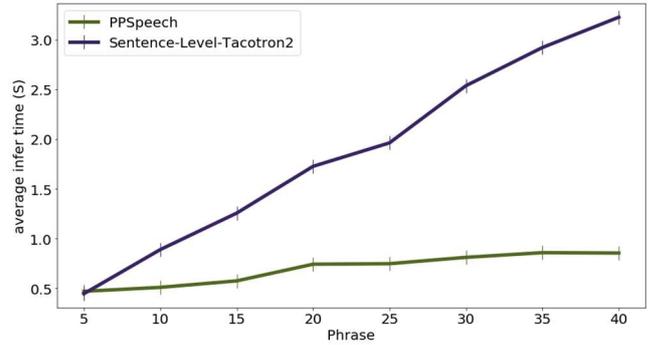

Figure 7: *Speed and Preference between Sentence-Level-Tacotron2 and PPSpeech.*

As shown in Figure 7, two systems has almost the same speed for 5-phrase sentences. It is because PPSpeech need to calculate additional context embedding and the phrase encoding of phone sequence may be implemented more than once due to the sliding window strategy. It can be optimized if the phrase encoding results are cached. As expected, with the growth of phrase number, the time cost of PPSpeech increase very slowly, while Sentence-Level-Tacotron2 rises remarkably. When a sentence contains 40 phrases, the time cost of Sentence-Level-Tacotron2 is about 5 times that of PPSpeech.

## 4. Conclusions

This work proposes PPSpeech, a phrase-level parallel speech synthesis system that is conditioned on both acoustic embedding and context embedding. Experiment results prove that PPSpeech can provide much faster speech synthesis service for long sentences. When a sentence contains 20 phrases, it can achieve about 3 times speed. At the same time, PPSpeech generate comparable or even slight better voice quality of speech than sentence level Tacotron 2. It should be attributed to the acoustic embedding and text context embedding. Acoustic embedding helps to control the timber and speaking style in a stable mode. While context embedding ensure the prosody of phrase depends on not only the content but also the context or position in the sentence.

Moreover, across the sentence, the timber or speaking style can be switched to match the scenario by changing the reference audio for acoustic embedding. For text context embedding, it also can benefit the prosody coherence between neighboring sentences. In summary, PPSpeech can be expected to achieve better quality in paragraph synthesis tasks. Potentially, the ideas of PPSpeech is promising to be scaled to other TTS system or generation tasks.